\documentstyle[pra,aps,multicol,eqsecnum,epsfig]{revtex}

\def\be{\begin{eqnarray}}
\def\ee{\end{eqnarray}}
\def\l{\langle}
\def\r{\rangle}

\newtheorem{definition}{Definition}[section]

\begin{document}
\title{ Dynamics of open quantum systems initially entangled 
with environment:\\  Beyond the Kraus representation}
\author{
Peter \v{S}telmachovi\v{c}${}^{1}$ and
Vladim\'{\i}r Bu\v{z}ek${}^{1,2}$,
}
\address{
${}^{1}$ Research Center for Quantum Information, 
Institute of Physics, Slovak Academy of Sciences,\\
D\'ubravsk\'a cesta 9, 842 28 Bratislava, Slovakia \\
${}^{2}$Faculty of Informatics, Masaryk University, Botanick\'a 68a,
602 00 Brno, Czech Republic
}

\date{19 June 2001}

\maketitle

\begin{abstract}
We present a general analysis of the role
of initial correlations between the open system and an environment
on quantum dynamics of the open system.  
\end{abstract}


\begin{multicols}{2}

\section{Introduction}
\label{sec1}
Proper understanding of quantum dynamics of open systems is a very 
important task in many areas of physics ranging from
quantum optics, to quantum information processing and to quantum
cosmology \cite{Davies}. 
In general, one can assume an interaction between  the open system 
denoted as $A$ with the environment $B$. This environment is a 
quantum system with the Hilbert space of an arbitrary dimension.
The whole  $A+B$ system evolves unitarily. In most of the studies
on dynamics of open systems it is assumed that the open system and 
its environment are at the initial moment of their joint evolution
factorized \cite{Davies,Preskill}, that is they are described by the
density operator of the form
\be
\rho_{AB}  =  \rho_A \otimes \rho_B 
\label{1.1}
\ee
where $\rho_A$ is the initial state of system $A$ and $\rho_B$ is the
initial state of the environment. While the initial state of the open system
$A$ may vary the initial state of the environment $B$ is considered to be
determined by external conditions.
In this context it is natural to ask what is the time evolution of
the open system $A$? Or in other words, what is the explicit form of the map
$ \$_A: \rho_A   \rightarrow  \rho_A^{'} $. In order to answer this
question one might follow the arguments presented in Ref.~\cite{Preskill}
and to find the explicit expression for the density operator $\rho'_A$:
\be
\rho'_A & = & {\rm Tr}_B \; (U_{AB} \rho_{AB} U^{\dagger}_{AB} ) 
\nonumber
\\
& = & \sum_{\mu} \langle \mu | U_{AB} \; \rho_A 
\otimes \left( \sum_{\nu} p_{\nu} | \nu \rangle \langle \nu | \right) 
U^{\dagger}_{AB} |\mu \rangle 
\nonumber
\\
& = & \sum_{\mu, \nu} 
\langle \mu | \sqrt{p_{\nu}}  U_{AB}| \nu \rangle 
\rho_A \langle \nu | \sqrt{p_{\nu}} U^{\dagger}_{AB} 
| \mu \rangle 
\nonumber
\\
& = & \sum_{\mu, \nu} M_{\mu \nu} \rho_A M^{\dagger}_{\mu \nu},
\label{1.2}
\ee
where 
\be
M_{\mu \nu}=
\langle \mu | \sqrt{p_{\nu}}  U_{AB}| \nu \rangle .
\label{1.3}
\ee
This is the well known Kraus representation \cite{Kraus} of a 
super-operator $\$_A$ which has been studied and used in the literature
broadly. On the other hand  dynamics of open system 
in which initial correlations
between the system and the environment is taken into account 
has not been analyzed in detail yet. Taking into account
recent interest in quantum entanglement within the context of quantum
information processing it is appropriate to study in detail the
role of (quantum) correlations on dynamics of open quantum systems.
Some particular aspects of this problem have been discussed 
earlier in Refs.~\cite{Romero,Hakim,Pechukas}. In particular, in
Ref.~\cite{Romero}
decoherence and initial correlations in quantum Brownian motion
have been analyzed, 
 while
in Refs.~\cite{Hakim}
a motion of a free particle coupled to a lineary dissipative 
environment has been studied. The conection between 
the initial correlations and the complete positivity 
has been discussed  in   Ref.~\cite{Pechukas}.

In the present paper we present a general analysis of the role
of initial correlations between the open system and an environment
on quantum dynamics of the open system.

\section{The role of initial correlations}
\label{sec2}
In this section we will 
investigate the evolution of an open system $A$ which is initially 
correlated with the environment $B$.
Let us denote by $\sigma_i$ the 
generators of the group $SU(N)$
\cite{footnote1} where $N$ is the dimension of the Hilbert
space of the system $A$. In addition we denote by 
$\tau_j$ the generators of the group $SU(M)$ where $M$ is the
dimension of the Hilbert space of the environment $B$. 
Using this notation 
the most general density matrix of the system $A$ and the environment $B$ 
reads as
\be
\rho_{AB} = \frac{1}{N M} \left 
( \openone_{AB} + \alpha_i \sigma_i \otimes \openone_B 
+ \beta_j \openone_A \otimes \tau_j + \gamma_{i j} \sigma_i 
\otimes \tau_j \right ) 
\label{2.1}
\ee
while the density operator of the open system $A$ is obtained via
``tracing'' over the environment
\be
\rho_A = {\rm Tr}_B \; (\rho_{AB}) = 
\frac{1}{N} ( \openone_A + \alpha_i \sigma_i) .
\label{2.2}
\ee
So let us assume that the state (\ref{2.1}) is the initial state
of the whole $A+B$ system which evolves according to the given 
unitary matrix $U_{AB}$. Can we describe in this case
the evolution of the subsystem $A$ in the form analogous to Eq.~(\ref{1.2})?
In order to answer the question we have to insert into the equation
\be
\rho'_A =  {\rm Tr}_B \; ( U_{AB} \: \rho_{AB} \: U^{\dagger}_{AB})
\label{2.3}
\ee
the expression (\ref{2.1}) for the density operator $\rho_{AB}$ which 
results in 
\be
\rho'_A 
& = & \sum_{\mu} \langle \mu |  U_{AB} \; \frac{1}{N} 
( \openone_A + \alpha_i \sigma_i) \otimes\frac{1}{M} 
( \openone_B + \beta_j \tau_j) U^{\dagger}_{AB} | \mu \rangle 
\nonumber
\\
& & + \sum_{\mu} \langle \mu | U_{AB} \; 
\frac{( \gamma_{i j} - \alpha_i \beta_j)}{NM} \sigma_i \otimes \tau_j  \;  
U^{\dagger}_{AB} | \mu \rangle 
\nonumber
\\
& = & \sum_{\mu} \langle \mu |  U_{AB} \; 
( \rho_A \otimes \rho_B ) \; U^{\dagger}_{AB} | \mu \rangle 
\nonumber
\\
& & + \sum_{\mu} \langle \mu | U_{AB} \; 
\frac{( \gamma_{i j} - \alpha_i \beta_j)}{NM} 
\sigma_i \otimes \tau_j  \;  U^{\dagger}_{AB} | \mu \rangle 
\label{2.4}
\ee
After rewriting $\gamma_{ij}$ as  
$\gamma_{ij} = N.M \: \gamma'_{ij} + \alpha_i \beta_j$
we obtain from Eq.~(\ref{2.4}) the expression
\be
\label{2.5}
\rho'_A & = & \sum_{\mu, \nu} M_{\mu \nu} \rho_A M^{\dagger}_{\mu \nu} 
\\ \nonumber 
& & 
+ \sum_{\mu} \langle \mu | U_{AB} \;  
\gamma'_{i j} \sigma_i \otimes \sigma_j  \;  U^{\dagger}_{AB} | \mu
\rangle, 
\ee
where the operators $M_{\mu \nu}$ are given by Eq.~(\ref{1.3}).
We see that the resulting density operator describing the open system
$A$ during the time evolution consists of two terms.
The first term corresponds to the 
standard Kraus representation with no initial correlations 
as discussed in Sec.~\ref{sec1}.
The second term in the right-hand side of Eq.(\ref{2.5}) 
depends {\em only}
on the correlation
parameters $\gamma'_{ij}$ which do not {\em explicitly}
 depend on the particular
choice of the initial state of the open system $A$ (see below). In other
words, these parameters cannot be determined by performing a local measurement
on the initial state of the system $A$.

This second term makes the equation (\ref{2.5}) 
{\em inhomogeneous though 
linear} - we will discuss this consequence of initial  correlations
between the system and the environment in the following section. 

{\em Example 1.}
We may regard $\gamma'_{ij}$ as additional parameters which together with
the initial state of the environment 
$\rho_B$ and the unitary operator $U_{AB}$ determine the time
evolution of the open system initially prepared in the state $\rho_A$.

To illustrate the possible significance of $\gamma'_{ij}$ we will
study  a simple model describing dynamics of two qubits (spin-1/2
particles).  In this model
one of the qubits ($A$) plays the role of the open system while the 
second qubit ($B$) plays the role of the environment.
Let the unitary evolution operator 
$U_{AB}$ acting on the joint system of these two qubits is given by the
expression
\be
U = e^{- i  H t} = \openone \cos t  - i  H  \sin t   \, ,
\label{2.6}
\ee
where $H$ is the Hamiltonian 
\be
 H = \sigma_x \otimes \frac{1}{2} 
( \openone - \sigma_z) +  \openone \otimes\frac{1}{2} ( \openone + \sigma_z) 
\, ,
\label{2.7}
\ee
with $\sigma_j$ being Pauli matrices. The interaction described by the
Hamiltonian (\ref{2.7}) corresponds to the well-known controlled-NOT gate
\cite{Preskill}. 

Let us consider two initial conditions 
$\rho_{AB}^{(1)}$ and $\rho_{AB}^{(2)}$
for the two-qubit state, which in the computer basis $\{|0\r,|1\r\}$ read
\be
\rho_{AB}^{(1)} &=& |\alpha|^{2} |00 \rangle \langle 00| 
+ |\beta|^{2} |11 \rangle \langle 11|  \, , 
\nonumber
\\
\rho_{AB}^{(2)} &=& (\alpha |00\rangle + \beta |11 \rangle) 
( \alpha^* \langle 00| +  \beta^* \langle11|).
\label{2.8}
\ee
Obviously, the qubits $A$ and $B$ in these two-qubit states
are in the same state, i.e.
\be
\rho_A^{(1)} &=& {\rm Tr}_B \left[\rho_{AB}^{(1)}\right] = 
{\rm Tr}_B \left[\rho_{AB}^{(2)}\right] = \rho_A^{(2)} \, ,
\nonumber
\\ 
\rho_B^{(1)} &=& {\rm Tr}_A \left[\rho_{AB}^{(1)}\right]  
= {\rm Tr}_A \left[\rho_{AB}^{(2)}\right]= \rho_B^{(2)} \, .
\label{2.9}
\ee
On the other hand the parameters
$\gamma'_{ij}$ in the two-qubit states (\ref{2.8}) are different.
Therefore, this simple model with identical states of subsystems but
different correlations will illuminate the role of the correlations
on dynamics of open quantum systems.

With the unitary evolution (\ref{2.6}) the two-qubit systems with the
two initial conditions (\ref{2.8}) evolve at time 
$t=\pi /2$ into states such that the system $A$ is described by the
two density operators
\be
\rho_A^{(1)} (t=\pi /2) & = & \frac{1}{2} (  \openone + \sigma_3 )  \, ,
\nonumber
\\
\rho_A^{(2)} (t=\pi /2) & = & 
\frac{1}{2} \left[  \openone + (|\alpha|^{2} - |\beta|^{2}) \sigma_3\right] \, ,
\label{2.10}
\ee
respectively. We stress here that the open system has been in both cases
in the same initial state, i.e.
 $\rho_A^{(1)}=\rho_A^{(2)}=|\alpha|^2|0\r\l 0|
+ |\beta|^2|1\r\l 1|$, the environment itself was in both cases initially
in the same state as well. But due to different initial correlations 
between the system and the environment
the open system has evolved 
into two different
states $\rho_A^{(1)}(\pi /2)$ and $\rho_A^{(2)}(\pi /2)$.

This example illustrates that the initial correlations between the
system and its environment may play important role in the dynamics
of open systems.
Moreover, in most of  physical situations such correlations are present
and therefore they have to be taken into account.

\section{Master equation}
\label{sec3}
As follows from our previous discussion both the state of the environment
and the initial correlations between the environment and the open system
play significant role in the dynamics of the open system.  
Therefore, in order   
to characterize  completely the evolution, 
it is necessary to determine (fix)
 the set of the parameters $\{ \beta_j \} $, i.e. 
 the state $\rho_B$ of the environment, and the
parameters  $\{ \gamma'_{ij} \} $ describing the correlations.
The  parameters $\alpha_i$, $\beta_j$ and $\gamma'_{ij}$ are {\em arbitrary}
 conditioned 
that the matrix $\rho_{AB}$ describe a real physical state of the system $AB$, 
that is,  
 it is a density matrix. Specifically,
 if we represent one particular choice of parameters 
$\{ \alpha_i, \beta_j, \gamma'_{ij} \}$ 
as a point in a $( N^2 \cdot M^2-1) $ dimensional 
space ${\mathbf R}^{(N^2 \cdot M^2 -1)}$, then the 
set of physically relevant parameters 
$\{ \alpha_i, \beta_j, \gamma'_{ij} \} $ form 
a convex subset ${\cal S}$  in the space ${\mathbf R}^{(N^2\cdot M^2 -1)}$. 
For example, in the case of $\alpha_i$ ( the same holds for $\beta_j$ 
and $\gamma'_{ij}$ ) there is only a subset  ${\cal O}_A$ in the space 
${\mathbf R}^{(N^2 -1)}$ from which we can chose the parameters 
$\{ \alpha_i \} $ so that the 
density matrix $\rho_{AB}$ is a reasonable density matrix. 
Moreover this subset 
${\cal O}_A$ depends on the choice of the remaining parameters 
$\{ \beta_j \} $ and $\{ \gamma'_{ij} \} $. In other words the 
subset  ${\cal O}_A$ is different 
for different choices of $\{ \beta_j \} $ and $\{ \gamma'_{ij} \}$.
 For example, if the system $A$ is in 
a pure state then the only possible initial
density matrix of the system $A$ and the environment must have the form
$\rho_{AB}= \rho_A \otimes \rho_B $, 
so that all $\gamma'_{ij}$ have to be zero. 
 Or equivalently if some of the 
parameters  $\gamma'_{ij}$  are not zero then the state $\rho_A$ can't 
be a pure state.

Sometimes it's very useful to describe the evolution of the open system in
a form of a master equation. In order to do so we
firstly rewrite the evolution (\ref{2.5}) in terms of 
the left-right super-operator acting on the density operator $\rho_A$
\be
\label{3.1}
\rho_A (t) = \widehat{{\cal T}} (t) \; \rho_A (0) + \xi (t), 
\ee
where $\xi (t)$ is the inhomogeneous term which has its origin in the
presence of initial correlations between the open system and the
environment, i.e. from Eq.~(\ref{2.5}) we have
\be
\nonumber
\xi(t) = 
\sum_{\mu i j } \langle \mu | U_{AB} \;
\gamma'_{i j} \sigma_i \otimes \sigma_j  \;  U^{\dagger}_{AB} | \mu
\rangle.
\ee
We stress once again that the operator $\xi(t)$ does not depend
{\em explicitly} 
on the initial state of the open system $A$, only the range
of possible values of correlations is determined by the choice
of $\rho_A$ and $\rho_B$ (see the discussion above).
As follows from Eq.~(\ref{2.5}) 
the left-right action of the super-operator $\widehat{{\cal T}} (t)$
is equal to the following normal action
\be
\nonumber
\widehat{{\cal T}} (t) \; \rho_A (0) =
\sum_{\mu, \nu} M_{\mu \nu} \rho_A (0) M^{\dagger}_{\mu \nu}. 
\ee

>From our previous comments it follows that the choice of the
initial correlations restricts a set of density operators $\rho_A$ for
which Eq.~(\ref{3.1}) can be used.
For instance, 
for pure states the term $\xi(t)$ is always zero.
Therefore, if we would use Eq.~(\ref{3.1}) with nonzero $\xi(t)$
for describing dynamics of an open system initially prepared in a pure
state, we might end up with a completely unphysical situation.
As discussed above this subset is determined by the condition, that
dynamics (\ref{1.2}) has a physical meaning. This restriction reflects
quantum nature of correlations between the system and the environment and
have to be taken into account in the derivation of dynamics of 
open quantum systems which are initially correlated with the environment.

 We have to keep in mind that
 there is always only a subset
${\cal O}_A$ of all density matrixes of the system $A$ for which the equation
(\ref{3.1}) with a given $\xi(t)$ 
is valid. If, for example, $\xi(t) = 0$ then the equation
(\ref{3.1}) is valid for all $\rho_A$ and ${\cal O}_A = {\cal S}_A$ where
${\cal S}_A$ is a set of all density matrixes of the system $A$. Unless
$\xi(t)$ equals to zero  ${\cal O}_A$ is a subset of $ {\cal S}_A$.

After this preliminary comments we derive the master equation
following the formalism presented in Ref.~\cite{Buzek}.
Differentiating Eq.~(\ref{3.1}) according to time we obtain 
\be
\label{3.2}
\frac{\partial}{\partial t} \rho_A (t)  = 
\frac{\partial}{\partial t} \widehat{{\cal T}} (t)  \; 
\rho_A (0) + \frac{\partial}{\partial t} \: \xi (t)
 \ee
When we substitute $\rho_A(0)$ which formally can be determined
with the help of Eq.~(\ref{3.1}) \cite{footnote}
into  (\ref{3.2}) we find 
\be
\frac{\partial}{\partial t} \rho_A (t) 
= \left ( \frac{\partial}{\partial t} 
\widehat{{\cal T}} (t) \right ) \; 
\frac{1}{ \widehat{{\cal T}} (t)}  
\left[ \rho_A (t) - \xi (t) \right]  + \frac{\partial}{\partial t} \: 
\xi (t)
\label{3.3}
\ee
If we introduce a notation for the Liouvillian super-operator
\be
\label{3.4}
\widehat{{\cal X}} =\left( \frac{\partial}{\partial t} \widehat{{\cal T}} (t) 
\right ) \; 
\frac{1}{ \widehat{{\cal T}} (t)}
\ee
then the master  equation can be rewritten in the following form
\be
\label{3.5}
\left ( \frac{\partial}{\partial t} - \widehat{{\cal X}} \right ) \;  
\left[ \rho_A (t) - \xi (t)\right] = 0
\ee

If  the initial correlations were zero, then the  master equation 
(\ref{3.5}) reduces to the well known form (see for instance,
Ref.~\cite{Buzek})
\be
\label{3.6}
\left ( \frac{\partial}{\partial t} - 
\widehat{{\cal X}} \right ) \;  \rho_A (t) = 0
\ee
where the operator $\widehat{{\cal X}}$ is the same as  in Eq.~(\ref{3.5}). 
Taking into account the fact, that $\xi(t)$ does not depend on the initial
state $\rho_A(0)$ we can introduce the operator
\be
{\cal F}(t) = 
\left ( \frac{\partial}{\partial t} - \widehat{{\cal X}} \right ) \; 
\xi (t)
\label{3.7}
\ee
and rewrite the master equation (\ref{3.5}) in an inhomogeneous form
\be
\label{3.8}
\left ( \frac{\partial}{\partial t} -
\widehat{{\cal X}} \right ) \;  \rho_A (t) = {\cal F}(t).
\ee 
We stress once again that the super-operator $\widehat{{\cal X}}$
depends only on the initial state of the environment $\rho_B$ and
the parameters of the unitary evolution $U_{AB}$,while the whole
information about the initial correlations between the open system
and the environment is in the operator ${\cal F}(t)$.

\section{Discussion}

Till now we have studied how  initial 
correlations between the  open system
and the environment can influence the time evolution of the open system.
We have found that these correlations play an important role which
cannot be neglected. 
In this section  we will investigate  properties of  super-operators
(evolutions) $\$_A$ acting on an open system which is a part of the
composite system (open system $+$ environment). It is assumed that
two parts of the composite system can be initially correlated.
The composite system is considered to be closed so that it evolves unitarily
according to a given unitary operator $U_{AB}$. In what follows we 
will assume this evolution of the ``Universe'' to be given.

Firstly we define the most general super-operator (evolution)
which originates from a given $U_{AB}$. 
\begin{definition}
\label{mos-gen-evo}
Let $A$ is the system of interest,
$B$ is the rest of the Universe (the environment), 
and $U_{AB}$ is a given unitary evolution on the whole system.
Let us consider a map 
\be
{\cal P}:\rho_A \rightarrow \rho_{AB}
\label{4.0}
\ee 
which means that for each   $\rho_A$ we choose 
one $\rho_{AB}$ from a set of   all possible density
matrices of the Universe such that
\be
{\rm Tr}_B ( \rho_{AB})  = \rho_A 
\label{4.1}
\ee
The {\bf super-operator}
 which describes the most general evolution of the system
$A$ is given by the expression
\be
\$ : \rho_A  \rightarrow  \rho_A^{'} 
\label{4.2}
\ee
\be
\rho_A^{'} \equiv  {\rm Tr}_B \left ( { U}_{AB} \rho_{AB} 
{ U}_{AB}^{\dagger} \right )
\label{4.3}
\ee
\end{definition}

The map ${\cal P}$ in Eq.(\ref{4.0}) is related to
the preparation  of the state $\rho_A$ of the system $A$. 
We note that while preparing the state $\rho_A$ of the
system $A$  the state of the Universe is changed as well.
That is, in any act of the preparation of the system $A$ we prepare
a state $\rho_{AB}$ rather then an isolated state $\rho_A$ of only the
system  $A$ without affecting the system $B$. For
this reason  $\rho_{AB}$ describes a (correlated) state of the open system
and the environment (Universe). Moreover, since the preparation is an act in
which a classical information is encoded into a quantum system the map
${\cal P}$ is not necessarily linear. Therefore the state
$\rho_{B}={\rm Tr}_A[\rho_{AB}]$ might depend (even in a nonlinear way)
on the state $\rho_A$. For instance we can imagine the map ${\cal P}$
of the form:  ${\cal P}(\rho_A)=\rho_A\otimes \rho_A$ 
which describe the action similar to quantum cloning which obviously is not
possible within the framework of linear quantum mechanics. But can easily
be performed at the level of preparation of quantum states. Analogously we
can imagine a map 
${\cal P}(\rho_A)=\rho_A\otimes \rho_A^T$, where $\rho^T$ is a transposed
state. Taking into account that $U_{AB}$ is fixed then the only ``freedom''
in controlling the dynamics is the choice of the map ${\cal P}$.

It is clear from the construction that the super-operator $\$ $ is a trace
preserving map and that the final operator $\rho_A^{'}$ is Hermitian and
positive, i.e., it is a valid density matrix.
In what follows we will study some aspects of the
 of the evolutions of the form \ref{mos-gen-evo}:

\paragraph{}
>From the definition \ref{mos-gen-evo} it follows that for a given
$U_{AB}$ and an arbitrary map ${\cal P}$ not all evolutions $\$_A$ can be
realized. On the contrary there exists $U_{AB}$ and ${\cal P}$ such that a
given $\$_A$ can be realized. To see this let us consider a following example.

{\em Example 2.}
Using the
scenario (\ref{4.3}) we can perform any map
$\$: \rho_A \rightarrow \rho'_A$ on a given (known) initial state
$\rho_A$ of the system $A$.
 Specifically, let  
 $ \$: \rho_A \rightarrow \rho'_A$ is a given map. 
We assume that the map ${\cal P}$ acting during the preparation of the
system $A$ is such that the composite system has been prepared in the state
$\rho_{AB}$ 
\be
\nonumber
 \rho_{AB} = \rho_A \otimes \rho_B 
\ee
such that $\rho_B=\rho'_A$
The unitary transformation which realizes the desired map is then taken to be
\be
\nonumber
{ U}_{AB} = \sum_{i,j} |i \rangle_A \langle j| 
\otimes | j \rangle_B \langle i | .
\ee
Obviously, there is 
nothing surprising here since if we know the
initial state of the system $\rho_A$ exactly then we can 
perform an arbitrary map on the system. In some sense this situation
corresponds to a classical physics when a complete knowledge 
about  the state
of the system is implicitly always assumed.
Knowing the initial state precisely we can perform
any map we want.

\paragraph{}
Till know we had not considered the linearity condition in association
with the evolution $\$_A$. As we have already commented the unitary
evolution $U_{AB}$ is by the definition linear, but the preparation
map ${\cal P}$ might be nonlinear. At this moment we can ask which
conditions on ${\cal P}$ has to be imposed so that $\$_A$ is linear.
In order to proceed we remind us the definition of the
linearity of the evolution $\$_A$. Namely, $\$_A$ is linear if
\be
 \$_A\left(\sum_j \lambda_j \rho_A^{(j)}\right) =  \sum_j 
\lambda_j \$_A \rho_A^{(j)}.
\label{4.3a}
\ee
Now it is clear that if ${\cal P}$ is linear, in a sense that
\be
{\cal P}\left(\sum_i \lambda_i \rho_A^{(i)}\right)=
\sum_i\lambda_i {\cal P}( \rho_A^{(i)})
\ee
then the evolution $\$_A$ is linear. The linearity
of ${\cal P}$ is a sufficient condition for the linearity of
$\$_A$. On the other hand it is not the necessary condition. We might
imagine a nonlinear map ${\cal P}$ such that $\$_A$ is linear. To understand
this we formally represent the evolution $\$_A$ as
$\$_A = {\rm Tr}_B U_{AB}$, where we use notation such that
$U_{AB}(\rho_{AB})= U_{AB}\rho_{AB} U^\dagger_{AB}$.
Then the linearity of $\$_A$ 
(\ref{4.3a}) can be expressed as
\be
\label{con-lin}
{\rm Tr}_B U_{AB} {\cal P} \left(\sum_i\lambda_i \rho_A^{(i)}\right)
&=&\sum_i \lambda_i {\rm Tr}_A U_{AB} {\cal P} (\rho_A^{(i)}) 
\\
\nonumber
&=& {\rm Tr}_B U_{AB} \left[\sum_i \lambda_i {\cal P} (\rho_A^{(i)}) \right]
\ee
>From this last equation it follows that if the map ${\cal P}$ is linear,
then $\$_A$ is linear as well. On the other hand from the
linearity of $\$_A$ does not follow that ${\cal P}$ is linear.
This is a consequence of the
property of the partial trace operation ${\rm Tr}_B$. Specifically, 
from the identity (\ref{con-lin}) the equality
\be
U_{AB} {\cal P}\left(\sum_i \lambda_i  \rho_A^{(i)} \right)  
= \sum_i \lambda_i U_{AB} {\cal P} (\rho_A^{(i)})
\ee 
does not follow.

\paragraph{}
Next we will consider consequences of 
another possible restriction on ${\cal P}$. Namely, let us consider
a rather frequent condition, that the state of the environment $\rho_B$
does not depend on the state of open system $\rho_A$. That is
\be
{\rm Tr}_A {\cal P}(\rho_A)= \rho_B = \mbox {\rm const}
\label{cond}
\ee
for all $\rho_A$.
If $\rho_A$ is pure, then under the condition 
(\ref{cond}) the map ${\cal P}$ is uniquely defined such that 
${\cal P}(\rho_{A})=\rho_A\otimes \rho_B$. 
On the other hand if $\rho_A$ is impure,
then under the condition (\ref{cond}) the map ${\cal P}$ 
might not be uniquely specified, 
i.e. correlation between $A$ and $B$ can play a role.

If the condition (\ref{cond}) is fulfilled and in {\em addition} the
evolution $\$_A$ in the definition \ref{mos-gen-evo} 
is linear, then the map ${\cal P}$ can be chosen such that
${\cal P}(\rho_A)=\rho_A\otimes \rho_B$ for all $\rho_A$.
But this means that the evolution
$\$_A$ can be represented in the Kraus representation
\cite{footnote3}. Consequently, this
map is completely positive \cite{Preskill}.

\acknowledgements
This work was supported in part
by the European Union  projects EQUIP and QUBITS under the contracts 
IST-1999-11053 and IST-1999-13021, respectively.
We thank M\'ario Ziman for many stimulating discussions and Lajos Diosi 
for helpful comments.

\end{multicols}


\begin{thebibliography}{}
\bibitem{footnote1}
To be mathematically correct  $\sigma_i$ are not
generators of the corresponding group but $\sigma_i$ multiplied by the
complex constant $i$. The complex constant is usually omitted and the
$\sigma_i$ alone are called generators. The same remark holds for
generators $\tau_j$.
\bibitem{footnote2}
Here we emphasize that we {\em can} choose the factorised 
density operator. It does not mean that the correlated
density operator cannot be used. For more discussion see the end of the
previous paragraph.
\bibitem{footnote3}
We note that in the standard approach to Kraus representation 
\cite{Preskill} the
sequence of arguments is different. It is assumed that $\$_A$ is linear,
trace preserving and CP (which guarantees the Kraus representation) then
there exist $U_{AB}$ and $\rho_{AB}=\rho_A\otimes\rho_B$ which realizes
$\$_A$.
\end{thebibliography}
\end{document}